\documentclass[a4paper,superscriptaddress,aps,onecolumn,floats,nofootinbib]{revtex4}
\usepackage{hyperref}
\usepackage{tcolorbox}
\usepackage{epsfig}
\usepackage{amsmath}
\usepackage{amssymb}
\usepackage{amsmath}
\usepackage{amsthm}
\usepackage[utf8]{inputenc}
\usepackage{graphicx}
\definecolor{arash}{rgb}{1.0,0.0,1.0}

\usepackage[top=2.5cm,right=2.5cm,bottom=2.2cm,left=2.5cm]{geometry}  
\usepackage{listings}

\definecolor{begie}{rgb}{0.9,0.9,0.95}
\lstnewenvironment{BASH}
{\lstset{language=bash,
				 numbers=left,           
		numbersep=10pt, 
		basicstyle=\ttfamily,
		numberstyle=\footnotesize\color{gray},
		tabsize=4,	 
		xleftmargin = 0.5cm,
		breaklines=true,
		escapeinside={(*@}{@*)}, 
		backgroundcolor=\color{begie},
		keepspaces=true,
		keywordstyle=\color{blue},
		frame=l
		 }}
{}
\lstnewenvironment{PYTHON}
{\lstset{language=Python,
				 numbers=left,           
				numbersep=10pt, 
				basicstyle=\ttfamily,
				numberstyle=\footnotesize\color{gray},
				tabsize=4,	 
				xleftmargin = 0.5cm,
				breaklines=true,
				escapeinside={(*@}{@*)}, 
				backgroundcolor=\color{begie},
				keepspaces=true,
				keywordstyle=\color{blue},
				frame=l
}}
{} 

\usepackage{placeins}

\let\Oldsection\section
\renewcommand{\section}{\FloatBarrier\Oldsection}

\let\Oldsubsection\subsection
\renewcommand{\subsection}{\FloatBarrier\Oldsubsection}

\let\Oldsubsubsection\subsubsection
\renewcommand{\subsubsection}{\FloatBarrier\Oldsubsubsection}

\begin{document}

\title{A Guide for CosmoMC Installation and Running}

\author{Hossein Moshafi}
\email{moshafi86@gmail.com}

\affiliation{Ibn-Sina Laboratory, Shahid Beheshti University, Velenjak, Tehran 19839, Iran}

\author{Majid Bahraminasr}
\email{majid.bahrami.nasr@gmail.com }

\affiliation{Ibn-Sina Laboratory, Shahid Beheshti University, Velenjak, Tehran 19839, Iran}

\begin{abstract}
CosmoMC is a Cosmological Monte Carlo package that explores parameter space, finds the best-fit values, and makes contour plots for various observational data. The present manual assists you with the installation steps and running of CosmoMC. Also, we briefly explain  Markov chains analysis and generating plots and tables for the parameters. This guide is for everyone who is not  familiar with GNU/Linux and wants to install and run CosmoMC for the first time.
\end{abstract}
\maketitle

\vspace{0.1cm}

\section{Introduction}
\label{sec:introduction}

CosmoMC ({``Cosmological Monte Carlo''}) is a free and open-source software that allows you to explore cosmological parameter space. It includes a Markov Chain Monte Carlo (MCMC) engine and some tools to analyze Markov chains and importance sampling. CosmoMC is written in FORTRAN 2003/2008. Also, There are some Python packages for building grids of runs, analyzing, and plotting. This code also includes CAMB\footnote{\url{http://camb.info}} code to calculate the theoretical matter power spectrum and CMB temperature and polarization power spectra. CosmoMC parallel implementation allows you to run the code on a multicore machine as well as  a massive cluster of thousands of cores. For more detailed documentation of this code, you can refer to: 
\url{https://cosmologist.info/cosmomc/readme.html}

\section{Installation and Preparation}

\subsection{Prerequisites}
\label{subsec:prerequisites}
\begin{itemize}
\item \textbf{Hardware} :  To employ the provided parallel capabilities of CosmoMC, you need more than just one core. Markov chain algorithms are time-consuming and performing them on parallel machines will boost the code's performance.

\item \textbf{Operating system}: CosmoMC is available for GNU/Linux distributions. Here, we installed and tested CosmoMC on \textbf{Ubuntu 18.04 LTS}\footnote{\url{http://releases.ubuntu.com/18.04/}}. Following the instruction's steps do not require prior knowledge of the GNU/Linux command line, however, it is highly recommended to see Appendix \ref{intro_2_gnu_linux} for a quick tour.

\item \textbf{Intel@FORTRAN Compiler (version 14+)}: CosmoMC is compatible with Intel Fortran. But You should use Intel Fortran versions older than 2019 version. CosmoMC does not compatible with the most recent version of Intel Fortran. We recommend to use \textbf{Intel Parallel Studio XE 2015} for the best performance.

\item \textbf{Open-MPI}: OpenMPI library is required for running on  parallel machines.

\item \textbf{CFITSIO}: CosmoMC uses the CFITSIO library to read file FITS data format including CMB data files.

\item \textbf{FFTW} and \textbf{GSL}: CosmoMC also uses 
\textbf{F}astest \textbf{F}ourier \textbf{T}ransform in the \textbf{W}est  (\textbf{FFTW}) and \textbf{G}NU \textbf{S}cientific \textbf{L}ibrary (\textbf{GSL}) to perform some actions and computations. Although their installation is not mandatory, It might be required in some operations.

\item \textbf{\emph{Planck} Likelihood Code and Data}:  \emph{Planck} Likelihood Code \emph{V3.0} is required to run CosmoMC with \emph{Planck 2018} data.

\end{itemize}
 
\subsection{Preparation}
\label{subsec:preparation}
In addition to the libraries mentioned earlier, there are some libraries and packages that may not include on a freshly-installed Ubuntu distribution. To make sure all the dependencies are satisfied, issue these commands in the terminal:

\begin{BASH}
 sudo apt update && sudo apt upgrade
 sudo apt install gfortran
 sudo apt install g++
 sudo apt install make
 sudo apt install gedit
 sudo apt install gnuplot
 sudo apt install gnuplot-x11
 sudo apt install python-pip
 sudo apt install cython
 sudo apt install python-numpy
 sudo apt install python-matplotlib
 sudo apt install python-scipy
 sudo apt install liblapack-dev
 sudo pip install pyfits
 sudo pip install getdist
\end{BASH}

\subsection{Installing FORTRAN Compiler}
\label{subsec:fortraninstall}

Download \textit{Intel Parallel Studio XE 2015 Update 1} from  \url{https://software.intel.com/en-us/fortran-compilers}. After extracting the compressed file, navigate into the extracted directory in terminal using \verb|cd| command and enter this command:
\begin{BASH}
 ./install_GUI.sh
\end{BASH}
Follow the installation wizard step by step. You may either specify the installation path or leave it with default path  \verb|/opt/intel/|. In order to make \verb|ifort| a system-wide command, you need to use this command:

\begin{BASH}
 source /opt/intel/composer_xe_2015.1.133/bin/ifortvars.sh intel64
\end{BASH}
 To define this command permanently, you have to add to this command at the end of \verb|.bashrc|, you may open \verb|bashrc| file with this command:
\begin{BASH}
 gedit $HOME/.bashrc &
\end{BASH}
Save changes and close the file. Then run this command in the terminal:
\begin{BASH}
 source $HOME/.bashrc
\end{BASH}
To check if the compiler works properly, run this command:
\begin{BASH}
 ifort -V
\end{BASH}
The output should be like this:
\begin{BASH}
 Intel(R) Fortran Intel(R) 64 Compiler XE for applications running on Intel(R) 64, Version 15.0.1.133 Build 20141023
\end{BASH}

\subsection{Installing Open MPI}
\label{subsec:openmpi}
In order to install Open-MPI, you need to compile it from the source code. Download Open-MPI from \url{https://www.open-mpi.org} , extract the compressed file and then navigate to the extracted directory. 
You need to specify  \emph{installation path}, \emph{FORTRAN compiler path}, \emph{FORTRAN 77 compiler path} and \emph{FORTRAN 90 compiler path}. In this stage you need to run configure command with appropriate flags which are presented below:
\begin{BASH}
 ./configure --prefix=/usr/local/ F77=/opt/intel/composer_xe_2015.1.133/bin/intel64/ifort FC=/opt/intel/composer_xe_2015.1.133/bin/intel64/ifort F90=/opt/intel/composer_xe_2015.1.133/bin/intel64/ifort
\end{BASH}
Then compile and install Open-MPI using the commands below:
\begin{BASH}
 sudo make
 sudo make install
\end{BASH}
The configuration and the installation process takes a rather long time. After the installation completed successfully, you need to add Open-MPI to the system path; open \verb|bashrc| file:
\begin{BASH}
 cd 
 gedit .bashrc &
\end{BASH}
and add these lines to \verb|bashrc| file:
\begin{BASH}
 PATH="/usr/local/bin":${PATH}
 export PATH
 LD_LIBRARY_PATH="/usr/local/lib/openmpi":$LD_LIBRARY_PATH
 export LD_LIBRARY_PATH
\end{BASH}
In some versions of Open MPI you may need to run following command in the terminal:
\begin{BASH}
 sudo ldconfig
\end{BASH}
To check if Open-MPI has been installed properly, you may run this command: 
\begin{BASH}
 mpif90 -V
\end{BASH}
And the output would be Intel FORTRAN compiler version.

\subsection{Installing CFITSIO}
\label{subsec:cfitsio}
 Download CFITSIO from \url{http://heasarc.gsfc.nasa.gov/FTP/software/fitsio/c/cfitsio_latest.tar.gz}. Extract and navigate to the extracted directory. Use the \verb|configure|, \verb|make|, \verb|make install| commands  to prepare, compile and install the compiled binary files :\\

\begin{BASH}
 tar zxvf cfitsio_latest.tar.gz
 cd cfitsio/
 sudo ./configure --prefix=/usr/local
 sudo make
 sudo make install
\end{BASH}

After the installation successfully finished, it is essential to add the installation directory to the system PATH: 
\begin{BASH}
 cd ~
 gedit .bashrc &
\end{BASH}
Add this line to the end of \verb|bashrc| file:
\begin{BASH}
 export LD_LIBRARY_PATH=/usr/local/lib:${LD_LIBRARY_PATH}
\end{BASH}
Save changes and run this command:
\begin{BASH}
 cd ~
 source .bashrc
\end{BASH}

\subsection{Installing FFTW \& GSL}
\label{subsec:fftw-gsl}
You can download FFTW and GSL from \url{https://www.gnu.org/software/gsl/} and \url{http://www.fftw.org/download.html}. the installation steps are similar to CFITSIO (see subsection \ref{subsec:cfitsio}).

\subsection{Building \emph{Planck} Likelihood Code}
\label{subsec:plc}
After downloading \textbf{PLC 3.01} from: \url{http://pla.esac.esa.int/pla/aio/product-action?COSMOLOGY.FILE_ID=COM_Likelihood_Code-v3.0_R3.01.tar.gz}, extract it and enter the extracted directory:
\begin{BASH}
 tar xzvf COM_Likelihood_Code-v3.0.R3.01.tar.gz
 cd  /code/plc_3.0/
\end{BASH}
In \verb|plc-3.0| there is a subdirectory named \verb|plc-3.01|, copy this directory to wherever you intend to be installed to, here the \verb|home| directory has been chosen, but feel free to choose any other directory.

In the \verb|Makefile| in \verb|plc-3.01| directory, compiler and library paths might need modification, open \verb|Makefile|:

\begin{BASH}
 gedit Makefile &
\end{BASH}
Find \verb|CFITSIOPATH| and check if its value match to the path of CFITSIO
\begin{BASH}
 CFITSIOPATH=/usr/local
\end{BASH}
Also, check the path of MKL :
\begin{BASH}
 MKLROOT=/opt/intel/composer_xe_2015.1.133/mkl
 LAPACKLIBPATHMKL= -L$(MKLROOT)/lib/intel64
\end{BASH}
Save and exit. Make sure the current directory is \verb|plc-3.01| and  run these commands:
\begin{BASH}
 ./waf configure --install_all_deps
 ./waf install
\end{BASH}
After the installation of \emph{Planck} code finished, you need to  create some environmental variables by adding these lines to the end of \verb|bashrc| file:

\begin{BASH}
 export PLANCKLIKE=cliklike
 export CLIK_PATH=$HOME/plc-3.01
 export LD_LIBRARY_PATH=$LD_LIBRARY_PATH:$CLIK_PATH/lib
\end{BASH}
The value of \verb|CLIK_PATH| is the absolute path for the \verb|plc-3.0|, if you have chosen another path you may specify it here. Then in the terminal we source \verb|bashrc| file again:
\begin{BASH}
 cd ~
 source .bashrc
\end{BASH}
It is essential to put \emph{Planck} likelihood data into \verb|plc-3.01| folder. We can download baseline data of \emph{Planck 2018} from \url{http://pla.esac.esa.int/pla/aio/product-action?COSMOLOGY.FILE_ID=COM_Likelihood_Data-baseline_R3.00.tar.gz}. Extract it, and copy \verb|hi_l| , \verb|low_l| and \verb|lensing| directory into the \verb|plc-3.01| directory.

\subsection{Compiling CosmoMC \& GetDist}
\label{subsec:cosmomc-getdist}
Now you ready to compile CosmoMC from the source code. You can download the latest version of this code from \url{https://cosmologist.info/cosmomc/submit.html} 
or clone the git repository by this command: 
\begin{BASH}
 git clone --recursive https://github.com/cmbant/CosmoMC.git
\end{BASH}
There is a \verb|data| subdirectory inside the \verb|CosmoMC| directory. if you need to use \emph{Planck 2018} data, It is essential to make a symbolic link to connect \verb|data| directory to \verb|plc-3.01|:
\begin{BASH}
 cd CosmoMC/
 ln -s $HOME/plc-3.01 ./data/clik_14.0
\end{BASH}
Now navigate to \verb|source| inside the CosmoMC directory and compile the source code using \verb|make| command 
\begin{BASH}
 cd source/
 make
\end{BASH}
After successful compiling, a file named \verb|cosmomc| will be created in \verb|CosmoMC| directory. 

\section{Running CAMB}
There is  subdirectory inside the \verb|CosmoMC| directory named \verb|camb|. CAMB\footnote{\url{https://camb.info}} is a Code for Anisotropies in the Microwave Background which is responsible for theoretical calculations of CosmoMC. For compiling CAMB, go into its directory and use \verb|make| command:
\begin{BASH}
 cd camb/fortran/
 make
\end{BASH}
The subdirectory \verb|inifiles| inside the \verb|camb| contains some input parameter files such as \verb|params.ini| or \verb|planck_2018.ini| . You may use these files as an input file for running CAMB. It worth mentioning that all these files can be edited. To use them as an input, you may copy one of these files in the \verb|camb| directory. After making CAMB Fortran files, it produces an executable file named \verb|camb| which can be run like this: 
\begin{BASH}
 ./camb params.ini
\end{BASH}
\begin{figure}[h]
\begin{center}
\includegraphics[scale=0.6]{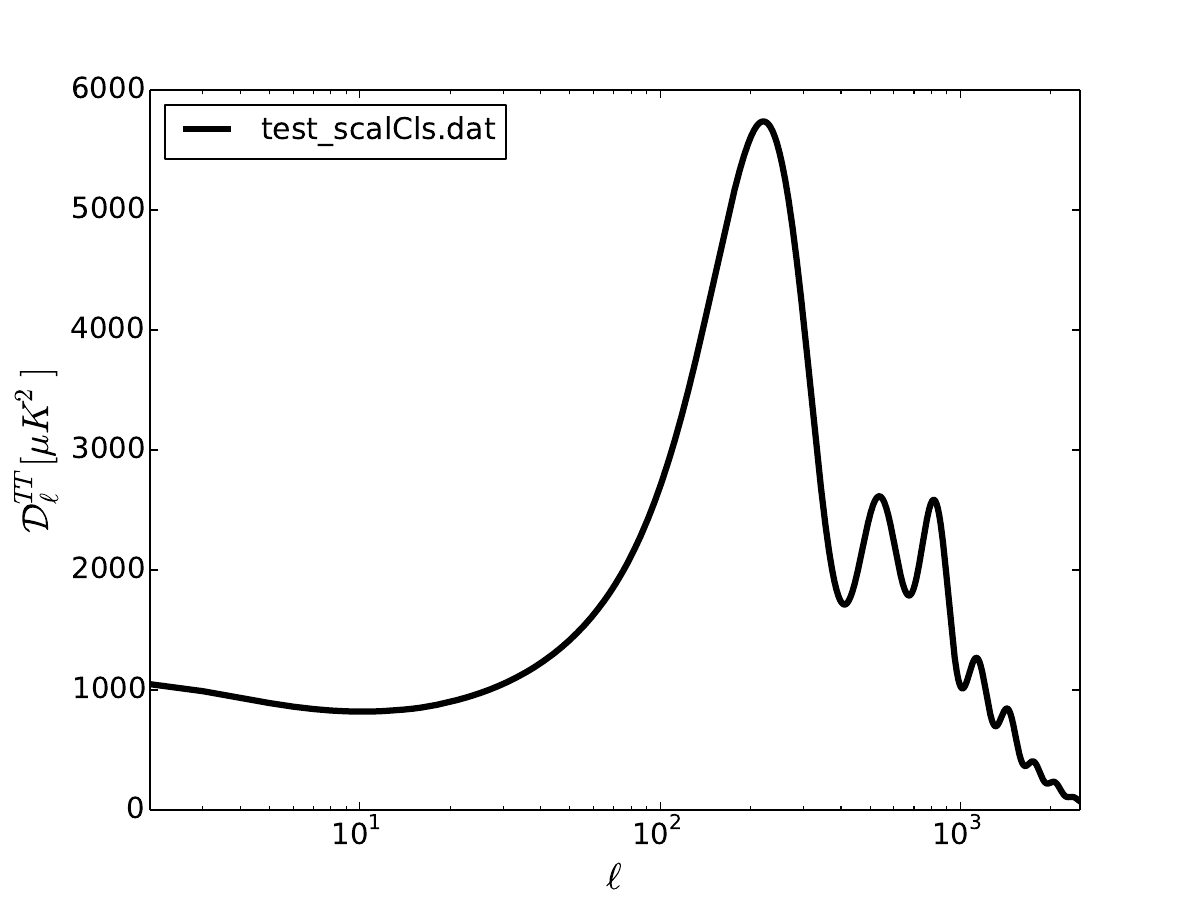}\caption{Scalar power spectrum as an output of CAMB code} \label{fig1}
\end{center}
\end{figure}
Outputs have \verb|*.dat| extension such as \verb|test_scalCls.dat| for scalar perturbations or \verb|test_tensCls.dat| for tensor perturbations power spectrum. We can plot them by using \verb|gnuplot| program. Run this command in terminal:
\begin{BASH}
 gnuplot
\end{BASH}
Now you may use the commands below to the plot power spectrum :
\begin{BASH}
 set logscale x
 plot"test_scalCls.dat" u 1:2 w l lw 2
 plot"test_scalCls.dat" u 1:3 w l lw 2
 replot"test2_scalCls.dat" u 1:2 w l lw 3
\end{BASH}
Output would be similar to Fig. \ref{fig1}.

\section{Running CosmoMC}
\label{sec:running}
To run CosmoMC you need to run it with MPI using \verb|mpirun| command. Set your corrent directory to \verb|CosmoMC| directory and run this command:
\begin{BASH}
 mpirun -np 7 ./cosmomc test.ini
\end{BASH}
The number \textbf{7} in front of the (\verb|-np|) flag specifies the number of the processors. If your machine has less than 7 processors you'll get an error. For checking number of processors , you may run this command:
\begin{BASH}
 cat /proc/cpuinfo
\end{BASH}
For running with \emph{Planck 2018} likelihoods you may use \verb|test_planck.ini|. However, it takes more time than \verb|test.ini| since it was just a test, not an actual MCMC run. 
For a real MCMC run, you may edit \verb|test.ini| file to change the setting:
\begin{BASH}
 gedit test.ini &
\end{BASH}
Inside the \verb|test.ini| file there are some lines which start with \verb|DEFAULT| keyword. If these lines are not  commented, it means we can use likelihoods related to these data. For example, if you want to include just \textbf{PlanckTT high-$\ell$ + PlanckTT low-$\ell$} your file should be similar to the following lines:

\begin{BASH}
#Bicep-Keck-Planck 2015, varying cosmological parameters (use only if varying r)
#DEFAULT(batch3/BK15.ini)

#DES 1-yr joint
#DEFAULT(batch3/DES.ini)

#Planck 2018 lensing (native code, does not require Planck likelihood code)
#DEFAULT(batch3/lensing.ini)

#BAO compilation
#DEFAULT(batch3/BAO.ini)

#Pantheon SN
#DEFAULT(batch3/Pantheon18.ini)

#Planck 2018, default just include native likelihoods (others require clik)

#high-L plik likelihood
DEFAULT(batch3/plik_rd12_HM_v22_TT.ini)

#low-L temperature
DEFAULT(batch3/lowl.ini) 

#low-L EE polarization
#DEFAULT(batch3/simall_EE.ini)

#general settings
DEFAULT(batch3/common.ini)

\end{BASH}
The line \verb|DEFAULT(batch3/plik_rd12_HM_v22_TT.ini)| is for including \textbf{PlanckTT high-$\ell$} and \verb|DEFAULT(batch3/lowl.ini)| is for \textbf{PlanckTT low-$\ell$} likelihood. Also, for \textbf{Planck 2018 CMB lensing} likelihood, you can use \verb|DEFAULT(batch3/lensing.ini)| line. For full description of \emph{Planck} data and likelihoods you may refer to 
\url{https://wiki.cosmos.esa.int/planck-legacy-archive/index.php/CMB_spectrum_%26_Likelihood_Code} .\\

To do a MCMC run you need to find the line which starts with \verb|action| and set its value to \textbf{0} :
\begin{tcolorbox}
\begin{BASH}
#action= 0 runs chains, 1 importance samples, 2 minimizes
#use action=4 just to quickly test likelihoods
\end{BASH}
\verb|action = 0| $ \Leftarrow$
\end{tcolorbox}

Also, you can turn on \textbf{checkpoint} option to build some checkpoints during running. Checkpoint lets you resume MCMC when your run is interrupted for any reason: 
\begin{BASH}
checkpoint = T
\end{BASH}
After making these changes in \verb|test.ini| file you may run it:

\begin{BASH}
 mpirun -np 7 ./cosmomc test.ini
\end{BASH}
Now it will start a MCMC run  to find the best-fit values and confidence intervals for your model.

\section{Analyzing Markov Chains}
\label{sec:analyzing-markov-chains}
After running of MCMC process finished, we can analyze chains to make the plots or table of the best-fit and confidence interval of parameters. For this purpose, we use a simple \textbf{Python} script. We have already installed a Python package called \textbf{GetDist} in Sec. \ref{subsec:prerequisites}.

\begin{figure}[h]
\begin{center}
\includegraphics[scale=0.4]{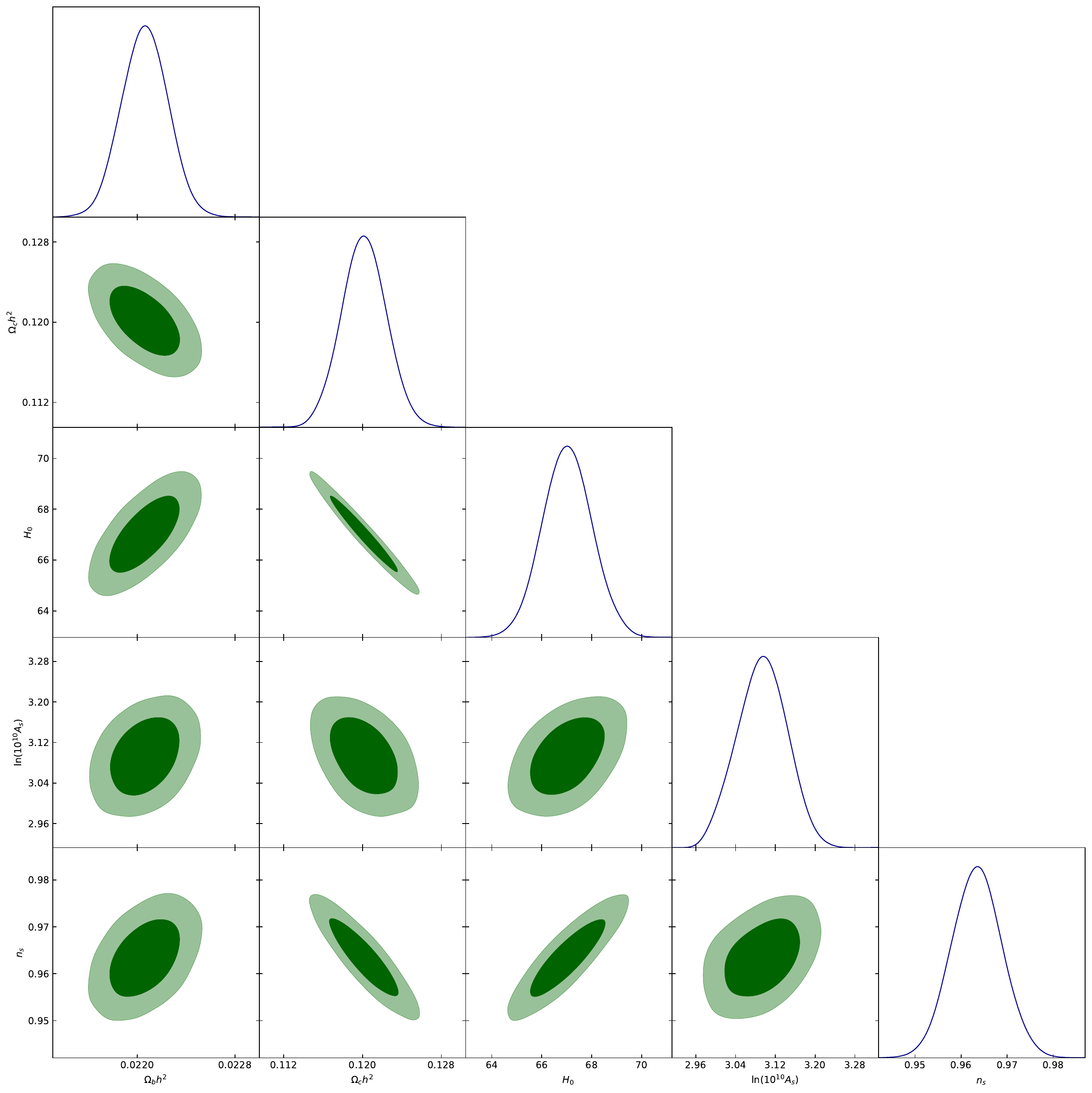}\caption{Contour plots and likelihoods as a product of GetDist Python package in CosmoMC} \label{fig2}
\end{center}
\end{figure}

Also, we can install Python packages included in CosmoMC from source files:
\begin{BASH}
 cd CosmoMC/python
 sudo python setup.py install
\end{BASH}
You can create triangular stack of likelihood contour plots using \verb|getdist| package in python:
\begin{BASH}
from getdist import plots, MCSamples
f=plots.getSubplotPlotter(chain_dir='/chains/',width_inch=12)
f.triangle_plot('cmb',['omegabh2','omegach2','H0','logA','ns'],filled=True,
legend_labels=['PlanckTT}'], legend_loc='upper right',
line_args=[{'lw':1, 'color':'darkblue'}], contour_colors=['darkgreen'])
f.export('mymodel_cmb_tri.pdf')
\end{BASH}
In this script \verb|/chains/| is the directory where contains the Markov chains, \verb|cmb| is the common name of the chains. For example, if we run CosmoMC on 10 processors with common name \verb|cmb| we will have chains with names of \verb|cmb_1.txt, cmb_2.txt , cmb_3.txt , ... cmb_10.txt|. Now you have to determine parameters which you want to plot in \verb|f.triangle_plot| function and also you may choose a legend for your plot. A sample output plot is represented in Fig. \ref{fig2}. \\
\\
For generating the table of parameters we can use this script:

\begin{PYTHON}
import getdist.plots as gplot
g = gplot.getSinglePlotter(chain_dir=r'/chains/')
samples = g.sampleAnalyser.samplesForRoot('cmb')
print(samples.getTable().tableTex())

print(samples.getInlineLatex('omegabh2',limit=1))
print(samples.getInlineLatex('omegach2',limit=2))

\end{PYTHON}
Set \verb|limit=1|, and  \verb|limit=2| to produce a \LaTeX-format table of parameters with $\%68$ and $\%95$ confidence level respectively.

\begin{center}
	\begin{tabular} { l  c}
		
		Parameter &  95\% limits\\
		\hline
		{\boldmath$\Omega_b h^2   $} & $0.02217^{+0.00027}_{-0.00026}$\\
		
		{\boldmath$\Omega_c h^2   $} & $0.1209^{+0.0026}_{-0.0026}$\\
		
		{\boldmath$100\theta_{MC} $} & $1.04084^{+0.00061}_{-0.00063}$\\
		
		{\boldmath$\tau           $} & $0.058^{+0.023}_{-0.018}   $\\
		
		{\boldmath${\rm{ln}}(10^{10} A_s)$} & $3.052^{+0.049}_{-0.035}   $\\
		
		{\boldmath$n_s            $} & $0.9580^{+0.0072}_{-0.0070}$\\
		
		$H_0                       $ & $66.8^{+1.1}_{-1.1}        $\\
		
		$\Omega_m                  $ & $0.322^{+0.017}_{-0.016}   $\\
		
		$\sigma_8                  $ & $0.816^{+0.022}_{-0.016}   $\\
		
		$z_{\rm re}                $ & $8.0^{+2.3}_{-1.8}         $\\
		
		$10^9 A_s                  $ & $2.12^{+0.11}_{-0.072}     $\\
		
		$Y_P                       $ & $0.24531^{+0.00011}_{-0.00012}$\\
		
		${\rm{Age}}/{\rm{Gyr}}     $ & $13.824^{+0.045}_{-0.045}  $\\
		
		$z_*                       $ & $1090.25^{+0.51}_{-0.51}   $\\
		
		$\chi^2_{\rm plik}         $ & $2441\,({\nu\rm{:}\,497.3})$\\
		
		$\chi^2_{\rm lowl}         $ & $25.2\,({\nu\rm{:}\,0.7})  $\\
		
		$\chi^2_{\rm CMB}          $ & $2466\,({\nu\rm{:}\,502.1})$\\
		\hline
		$\Omega_b h^2          $ &  $0.02217\pm 0.0001$\\
		$\Omega_c h^2           $ & $0.1209\pm 0.0013$\\
	\end{tabular}
\end{center}

For complete documentation of GetDist package you can refer to \url{http://getdist.readthedocs.io}.

\appendix
\section{Linux}
\label{intro_2_gnu_linux}
\subsection{GNU/Linux}
GNU/Linux or casually ``Linux" operating system, consists of a Linux kernel and some  GNU utilities and tools coded by or for the GNU project. GNU license make it possible for companies and individuals to make their own Linux distributions for various intents and purposes by adding additional software and documentation.  Here, Among the myriad of GNU/Linux operating systems, we used a Debian-based distribution name \textbf{Ubuntu}\footnote{\url{https://www.ubuntu.com}} for its active community and ease of use.  If you are not familiar with Linux installation, you may need to read this guide 
\url{https://www.wikihow.com/Install-Ubuntu-Linux}

\subsection{Command line Interface}

A GNU/Linux shell is a command-line interpreter. Here we used the default Ubuntu shell, Bash. But you feel free to use your favorite shell.

You can run bash commands in {\it Ubuntu Terminal} to open a terminal window, you may press \verb|ctrl+Alt+t| or click on the terminal icon. Also, there is an \verb|open in terminal| option on right-click menu that helps you to open a terminal and set Bash current directory to a specific directory without navigating to it by using command-line commands. 

Bash prompt has this structure
\begin{BASH}
 user_name@computer_name:address $ 
\end{BASH}
The flickering cursor in front of \$ sign means Bash is ready to get the commands. When you open terminal in the home directory, you do not see the full address as is described in \ref{Appendix-linux-address} but a ~ sign, which is an abbreviation for home directory address.

\subsubsection{Environment Variables}
The environment variable is a variable with a name and an associated value. In this manual, we mostly deal with \$USER, \$HOME, \$PATH variables, which their values are respectively, user's name and home directory address, and the system's path directories. Using echo command, you may see their values:
\begin{BASH}
 echo $HOME
 echo $USER
 echo $PATH
\end{BASH}
\subsubsection{Commands}
There are many Bash commands to do various operations and actions. Here, we introduce some commands which help to install and run CosmoMC

\begin{itemize}
\item {\bf pwd}: {\bf p}rint {\bf w}orking {\bf d}irectory  command.

\item {\bf cd}: This command helps you to change the working directory using relative or absolute address (see \ref{Appendix-linux-address})  

\begin{BASH}
 cd                 #returns to home directory
 cd ~               #goes to home directory
 cd $HOME           #goes to home directory
 cd x/              #goes to x directory using its reletive path
 cd ./x/            #goes to x directory using its reletive path
 cd /x/y/          #goes to y directory using its absolute path
\end{BASH}

\item {\bf sudo}: {\bf s}uper {\bf u}ser {\bf d}o; some actions require more than user privileges, this command gives root privileges to the user to accomplish the task.

\item {\bf ls}: This command list all the directories and files.

\item {\bf tar}: This command with specific options is used for extraction, for example the command below extract \verb|cfitsio_latest.tar.gz| tar archive:
\begin{BASH}
 tar zxvf cfitsio_latest.tar.gz
\end{BASH}
Also, you may extract the file by its designated right-click option.
\end{itemize}
\subsubsection{Installing new programs}
Installing new packages is possible either using a package manager and binary file or compiling from the source code. Ubuntu uses \verb|apt| package manager by default which needs root privileges:
\begin{BASH}
 sudo apt install gfortran
\end{BASH}    

Also, it is possible to compile from the source code which usually done by three steps:
\begin{itemize}
	\item {\bf configure}:The configure script used to set correct values for the compiling step. Also, it is possible user sets some values with special flags.
	\item {\bf make}: This command compiles the package according to the instruction in \verb|Makefile|.
	\item {\bf make install}:  This command installs the compiled files created in the last stage. It worth mentioning that sometimes you need root privileges to install the package.
\end{itemize} 
\subsubsection{GNU/Linux PATH}
PATH is an environmental variable that tells bash which directories to lookup when users issue a command. To see the list of these directories, enter this command in the terminal:
\begin{BASH}
 echo $PATH
\end{BASH}
Any other command/excutable files that are nAny other command/executable files that are not to the PATH should be either 
issued with its designated address, i.e.:
\begin{BASH}
 ./my_custom_command.sh
 $HOME/Desktop/my_custom_command.sh
\end{BASH}
or it should be exported to the Linux PATH, i.e.:
\begin{BASH}
 PATH="/ADDRESS/TO/CUSTOM_COMMAND":${PATH}
 export PATH
\end{BASH}
This action is temporary. In order to set them permanently, it should be added to \verb|.bashrc| file in the home directory. First open \verb|.bashrc| file by this command:
\begin{BASH}
 cd
 gedit .bashrc &
\end{BASH}
Export the address to PATH by adding these lines at the end of the file:
\begin{BASH}
 PATH="/ADDRESS/TO/CUSTOM_COMMAND":${PATH}
 export PATH
\end{BASH}
Save and close the file. Then logout/login or simply enter this command in terminal:
\begin{BASH}
 source  .bashrc &
\end{BASH}

\subsubsection{Filesystem Hierarchy Standard}
\label{Appendix-linux-address}
All the files and directories are contained under the root directory ``\verb|/|''. Every user has a subdirectory under the path ``\verb|/home/|''. Above this home directory designated to the user,  it is not possible to modify any file or directory without root privileges.

Every file and directory has its unique address, it can be specified from the top-most directory in a hierarchy, i.e. the root directory or starting from the current working directory. The first called absolute address and the latter called relative address. Here are some examples for navigation to the Desktop directory from the home directory using either absolute and relative address:
\begin{BASH}
 cd /home/$USER/Desktop/              #absolute path
 cd $HOME/Desktop/             #goes to home directory
 cd Desktop/          #goes to home directory
 cd ./Desktop
\end{BASH}
The address of file can be uniquely specified by this name and its directory address, for example, the address of \verb|camb| file on the \verb|Desktop| directory has the absolute address of \verb|/home/$USER/Desktop/camb|, the relative address has the same logic but it depends on the directory that you are in.

There is a subtle point about running the executable files; even if the file is at your current working directory you cannot run it like this:
\begin{BASH}
 camb          
\end{BASH}
Since Bash assumes every input as a command and will search the system PATH. So, if you have not exported your current directory to system PATH you face \verb|command not found| error. To avoid this situation issue the command like this:
\begin{BASH}
 ./camb          
\end{BASH}
In the Bash ``\verb|.|'' means ``here'', so you are telling Bash you are attempting to run \verb|camb| executable file which is presented in the current directory
\end{document}